\documentstyle[preprint,aps]{revtex}
\font\elevenbf=cmbx10 scaled\magstep 1
\begin{document}
\preprint{PRL-TH-94/18}
\title
{\bf ASTROPHYSICAL LIMITS ON GRAVITINO MASS\\}
\author{ Marek Nowakowski \footnote{Feodor Lynen Fellow}
\footnote{Address after $1^{st}$ July:
Laboratori Nazionali di Frascati, C.P. 13-Frascati (Rome), Italy.}
and Saurabh D. Rindani}
\address { Theory Group, Physical Research Laboratory,
Navrangpura, Ahmedabad 380 009, India}
\maketitle
\begin{abstract}
We calculate exotic cooling rates of stars due to
photo-production of light particles (mostly scalars and
pseudoscalars) which originate from the hidden sector of no
scale supergravity theories. Using this we can restrict the
gravitino mass $m_{3/2}$. The range of eliminated values of
$m_{3/2}$ stretches over six orders of magnitude and is given by
$\displaystyle{2 \times 10^8 <
{m_{\tilde g} \over m_{3/2}} < 6 \times 10^{13}}$,
$m_{\tilde g}$ being the gluino mass.
Combining our result with the earlier analysis from colliders ($\displaystyle{
{m_{\tilde g} \over m_{3/2}} < 2.7 \times 10^{14}}$) we conclude
that $\displaystyle{{m_{\tilde g} \over m_{3/2}} < {\cal
O}(10^8)}$ except for a narrow window around $10^{14}$.
Together with the current experimental limit on $m_{\tilde{g}}$
and cosmological constraints on $m_{3/2}$, albeit model
dependent, our analysis shows that a light gravitino is on the
verge of being ruled out.
\end{abstract}
\newpage
In the last twenty years a lot of effort went into establishing a
signal of a theory which we believe would replace the standard
$SU(3)_C \otimes SU(2)_L \otimes U(1)_Y$ model at high energies.
In the order of the increasing scale these theories are likely
to be: global supersymmetry \cite{susy}, supersymmetric Grand Unified Theory
(GUT) \cite{gut} and supergravity \cite{sugra} which,
being non-renormalizable, should
in principle be only an approximate version of something more
fundamental, like superstrings. Whereas we hope that the effects
of global supersymmetry will manifest themselves around $1 \;
TeV$ scale the physical effects of GUT theories which are
testable at present in laboratory experiments are only a few
(to mention the standard example of proton decay or the
existence of a monopole). Up to now searches for supersymmetric
particles and proton decay have resulted in lower bounds on
their masses and an upper bound on proton lifetime \cite{pdb}. The presence
of supergravity theory at Planck scale justifies the soft
supersymmetry breaking terms in the global theory i.e.
effectively reduces the number of free parameters which via
renormalization group equation can be scaled down to present
energies. A more direct signal of the usual supergravity is
practically non-existent. Though, in general, not a serious
problem, this can be a bit intriguing as supergravity theories
are not unique due to the arbitrariness of the K\"ahler potential.
It has been, however, realized that a superlight gravitino
($\tilde G$) could be the harbinger of physics from the hidden
sector of supergravity \cite{fayet1}. A superlight gravitino field $\psi_{\mu}$
acts as spin-1/2 Goldstino $\chi$ with $\psi_{\mu}=i \sqrt{2/3} m_{3/2}^{-1}
\partial_{\mu}\chi$. In momentum space the spin summed density matrix
corresponding to this longitudinal part of $\psi_{\mu}$ takes
the form $p_{\mu}p_{\nu}/m_{3/2}^2$ which effectively, when
coupled to a gauge boson $V$ ($V=\gamma,\;g,\; Z$) and a gaugino
$\tilde V$, enhances the coupling by a factor
$\displaystyle{{m_{\tilde V} \over m_{3/2}}}$. More precisely,
since $\tilde G$ is the superpartner of graviton it  couples to
particles with the strength proportional to $\kappa=\sqrt{8\pi G_N}=
4.11 \times 10^{-19}\;GeV^{-1}$ ($G_N$ is Newton's gravitational
constant). Hence the effective net coupling turns out to be
$\kappa m_{\tilde V} / m_{3/2}$. Later the idea of a
superlight gravitino was embedded in the no scale
supergravity theories \cite{ellis1}. Special attention was drawn to the
super-Higgs mechanism in these type of theories and it was
found that the coupling of certain light scalar
(${\cal S}$) and pseudoscalar (${\cal P}$) to gauge bosons is also
proportional $\kappa (m_{\tilde V} / m_{3/2})$ \cite{probir1}. These particles
(${\cal S}$ and ${\cal P}$), in the physical spectrum of the
theory, are members of a chiral superfield in the hidden sector.
Their spin-1/2 superpartners are Goldstinos which get eaten up by the
gravitino. Needless to say that a signal originating
from the hidden sector of such a supergravity theory would be quite
spectacular,
more so as the relevant coupling is directly proportional to the gravitational
constant $G_N$. Indeed bounds (to be discussed later) on
$m_{3/2}$ have been put by examining exotic decay processes and
scattering reactions involving ${\cal S}$, ${\cal P}$ and
$\tilde G$ at collider energies \cite{fayet2,probir2,nandi,dicus}.
It is worth mentioning here
that what makes bounds on gravitino mass even more interesting
are hints from cosmology that point in the direction of either a
light gravitino with mass $m_{3/2} < 1\; keV$ or a heavy one with
$m_{3/2} > 1 \; TeV$ \cite{primack,weinberg}. In this letter we
eliminate the parameter
space of $m_{3/2}$ further by investigating exotic cooling
processes of stars like the Sun and Red Giants. In particular
these additional cooling rates are due to the photon reactions like
$\gamma e^{\pm} \to \gamma+ {\cal S}/{\cal P}$ (Primakoff process)
and $\gamma \gamma \to \tilde{G}
\tilde{G},\; {\cal SS}, \; {\cal PP}$.

The interaction lagrangian relevant for single and two photon processes
is \cite{cremmer,probir1}
\begin{eqnarray}\label{e1}
e^{-1}{\cal L}_{int}&=& -{\kappa \over 4}\sqrt{{2 \over
3}}\biggl({m_{\tilde{\gamma}} \over m_{3/2}}\biggr)\biggl[{\cal
S}{\cal F}^{\mu \nu}{\cal F}_{\mu \nu}\;\; + \;\; {1 \over
2}{\cal P}\varepsilon^{\mu \nu \rho \sigma}{\cal F}_{\mu
\nu}{\cal F}_{\rho \sigma}\biggr] \nonumber \\
&+& {1 \over 2}\varepsilon^{\mu \nu \rho \sigma} \bar{\psi}_{\mu}\gamma_5
\gamma_{\nu}{\cal D}_{\rho}\psi_{\sigma}\;\; + \;\; {1 \over
4}\kappa \bar{\lambda}
\gamma^{\sigma}\sigma^{\mu \nu}\psi_{\sigma}{\cal F}_{\mu \nu}
\nonumber \\
&-&i{\kappa \over 2}\sqrt{{3 \over 2}}\biggl[m_{3/2}{\cal S}\bar{\psi}_{\mu}
\sigma^{\mu \nu}\psi_{\nu}\;\; + \;\; {1 \over 2}
\varepsilon^{\mu \nu \rho \sigma}\bar{\psi}_{\mu}\gamma_{\nu}\psi_{\rho}
\partial_{\sigma}{\cal P}\biggr] \end{eqnarray}
where $e$ is the determinant of the vierbein and ${\cal
D}_{\mu}$ is the covariant derivative expressed in terms of the
spin connection $\omega_{\mu ab}$ viz. ${\cal D}_{\mu}\psi_{\nu}
=(\partial_{\mu}-i/4 \omega_{\mu ab}\sigma^{ab})\psi_{\nu}$.
$\lambda$ is the photino field and $m_{\tilde{\gamma}}$ is the
mass parameter which enters the full lagrangian in the form of
the bilinear $m_{\tilde{\gamma}}\bar{\lambda}\lambda$. ${\cal
F}_{\mu \nu}$ is the usual electromagnetic stress tensor.
In general, one can have more than one pair of ${\cal S}$ and
${\cal P}$ particles, but for practical purposes it is
convenient to take just one pair.

The first two terms in eq.(1) are sufficient to calculate
$\gamma e \to \gamma + {\cal S}/{\cal P}$. The rest of the interaction
terms appears in the calculation of $\gamma \gamma \to \tilde{G}
\tilde{G}$ where we have two diagrams with $\tilde{\gamma}$ in
$t$- and $u$-channel as well as three s-channel diagrams with
$G$, ${\cal S}$, ${\cal P}$ in the intermediate state. Note also
that since the dimension $5$ operators ${\cal S}{\cal F} \cdot
{\cal F}$ and
${\cal P} {\cal F}\cdot \tilde{{\cal F}}$ contain higher order derivatives they
are expected to violate tree level unitarity at some scale. A
number of papers \cite{probir1,probir2} have been devoted to this subject.
The violation of tree level unitarity
need not be a serious embarrassment of the theory since
supergravity should not be considered as the last word in the
physics of Planck scale. We view it rather as an approximation
to the latter. Therefore as long as we do not stretch the
involved energies beyond a critical value
(in astrophysical applications this is never the case) the
results are still trustable.

At tree level ${\cal S}$ and ${\cal P}$ are strictly massless.
They can, however, acquire mass radiatively, mostly through
loops with intermediate gluons. Their mass is estimated to be \cite{ellis2}
\begin{equation} \label{e2}
m_{{\cal S}/{\cal P}}\;\; \sim \;\; \kappa \left({m_{\tilde{g}} \over
m_{3/2}}\right)\; \Lambda_{QCD}^2 \end{equation}
Commonly one assumes the equality of gaugino masses at a high
unification scale. Via renormalization group equation this
assumption leads to the following mass relation at lower
energies \cite{nilles}
\begin{equation} \label{e3}
m_{\tilde{\gamma}}\;\; \simeq \;\; {8 \over 3}\; {\alpha_{em} \over
\alpha_s} \; m_{\tilde{g}}\;\; \simeq \;\; {1 \over 6}\; m_{\tilde{g}}
\end{equation}
In what follows we will use eq.(3) as a guiding relation between
$m_{\tilde{g}}$ and $m_{\tilde{\gamma}}$. We mention this
explicitly as in view of eqs.(2) and (3) any bound on $\displaystyle{
\left({m_{\tilde{\gamma}} \over m_{3/2}}\right)}$ can be
translated into a bound on $m_{{\cal S}/{\cal P}}$ and vice
versa. Furthermore, the coupling $g_{{\cal S}/{\cal P}\gamma \gamma}$
depends now linearly on $m_{{\cal S}/{\cal P}}$ in which case  astrophysical
limits on $m_{{\cal S}/{\cal P}}$ and $g_{{\cal S}/{\cal P}\gamma \gamma}$
should not be treated independently, a situation encountered
also in axion physics.

The best model-independent bound on $\displaystyle{\left({m_{\tilde{\gamma}}
\over m_{3/2}}\right)}$ obtained so far comes from $e^+e^-$
colliders in the reaction $e^+e^- \to \gamma + {\rm nothing}$
realized in the model by $e^+e^- \to \gamma {\cal S},\;\; \gamma
{\cal P}$ \cite{dicus}. A slightly model dependent bound, assuming
$m_{\tilde{q}} > m_{\tilde{g}}$, can be derived in $p\bar{p}$
reactions studying the events of monojets and dijets with
missing $p_T$ \cite{nandi}. This
would be the signal of processes like $p\bar{p} \to g{\cal
S},\;\; g{\cal P},\;\;\tilde{g}
\tilde{g},\;\;\tilde{g}\tilde{G}$. The outcome of the two
analyses yields
\begin{equation} \label{e4}
\left({m_{\tilde{\gamma}} \over m_{3/2}}\right)_{e^+e^-}\;\; < \;\;
1.8 \times 10^{15}\;\;\;\; , \;\;\;\; \left({m_{\tilde{\gamma}} \over
m_{3/2}}\right)_{p\bar{p}}\;\; < \;\; 4.5 \times 10^{13} \end{equation}
corresponding to $m_{{\cal S}/{\cal P}}=0.18\;MeV$ and $m_{{\cal
S}/{\cal P}}=4.4\;keV$, respectively ($\Lambda_{QCD}=200\;MeV$).
Both limits are better than the older Fayet's bound of $4.3 \times
10^{15}$.

The importance of restricting gravitino mass becomes also clear
in light of the cosmological arguments given in \cite{primack,weinberg}.
Based on
the observed bound on cosmological mass density and a Helium
synthesis there are strong indications, quite independent of the
specifics of an underlying supergravity theory, that
\begin{equation} \label{e5}
1\; keV\;\;\; \vert \!\!\!\!\! <\;\;\;m_{3/2}\;\;\;
\vert\!\!\!\!\! <\;\;\; 10\;TeV
\end{equation}
i.e. cosmology seems to favour either a light or a heavy gravitino.
This restriction can be evaded by assuming an inflationary
scenario with low reheating temperature \cite{ellis3}. This is, however,
incompatible with baryogenesis from GUT theories which should
occur at higher temperature \cite{linde}. A natural way out would be to
produce baryon asymmetry through anomalous processes at
electroweak scale \cite{cohen}. Whether this is possible remains still an open
question, but there are convincing arguments which show that the
Higgs mass required for such scenario is in contradiction with
recent LEP limits on $M_{Higgs}$ \cite{cohen}. The baryon asymmetry  produced
at GUT scale with $B-L$ consevation
gets washed out by the anomalous processes unless e.g.
some lepton number violating processes are operative \cite{kuzmin}.
An alternative would be a GUT theory with no $B-L$ conservation
where the baryon asymmetry can, in general, survive the
electroweak transition.

Yet another way to avoid restriction (5) can be broadly
described as belonging to the class of R-parity breaking
theories \cite{rparity}. In any case, we will envisage a change of our
understanding of supergravity theories and/or cosmology, should
the limit given in eq.(5) be violated.

Before applying astrophysical processes \cite{astro} to restrict the
gravitino mass let us make some general comments on what is to be
expected in such an analysis. From the core temperature of the stars
like Sun ($T_{\odot} \simeq 1\; keV$), Red Giants ($T_{RG}
\simeq 10\; keV$), White Dwarfs ($T_{WD} \simeq 2\; keV$)
\cite{note1} and
a newly born Neutron Star in a Supernova explosion \cite{bethe} ($T_{SN} \simeq
30\; MeV$) it becomes evident that only light particles (in the
extreme case with a few $MeV$ mass) can contribute to the
cooling processes of stars. Plasmon effects will suppress single
and two gamma processes (relevant in our case) in White Dwarfs
and Supernovae. Hence we are left with the Sun and the Red
Giants where the available energy is at most few $keV$.
Moreover, if the particles ${\cal S}$ and ${\cal P}$ become too
heavy their relatively fast decay will preclude the possibility
of their causing any exotic cooling. From the width
\begin{equation} \label{e6}
\Gamma ({\cal S}/{\cal P} \to \gamma \gamma )= {1 \over 96 \pi}\kappa^2
\left({m_{\tilde{\gamma}} \over m_{3/2}}\right)^2\; m_{{\cal S}/{\cal P}}^3\;\;
\sim \;\; {1 \over 3456 \pi}\left({m_{{\cal S}/{\cal P}} \over \Lambda_{QCD}}
\right)^4 \; m_{{\cal S}/{\cal P}} \end{equation}
we can estimate the distance $\lambda_{flight}$ traveled by the
particle in the rest frame of the star in meters
\begin{equation} \label{e7}
\lambda_{flight} \simeq 3 \times 10^{15}\; \left({keV \over
m_{{\cal S}/{\cal P}}}\right)^6 \left({\omega' \over keV}\right)
\left(1- \left({m_{{\cal S}/{\cal P}} \over \omega'}\right)^2 \right)^{1/2}\;\;
m \end{equation}
where $\omega'$ is the energy of the produced particle. Let us apply
eq. (7) to the Primakoff process on a target $Z=e,\; p $ etc. In
the case of interest where the initial photon energy
$\omega_{\gamma}$ is $T_{star}$ we have $m_{Z} >>
\omega_{\gamma} >> M_{{\cal S}/{\cal P}}$. The four momentum transfer
$q$ is then very low, $\displaystyle{m_{{\cal S}/{\cal P}}^2
\left({m_{{\cal S}/{\cal P}}^2 \over 4 \omega^2_{\gamma}}\right)
\leq \vert q^2 \vert \leq 4 \omega^2_{\gamma}}$, such that
$\omega_{\gamma} \sim T_{star} \sim \omega'$. If we take
$m_{{\cal S}/{\cal P}} \sim 10^{-1}T_{star}$ we get for the Red Giants
$\lambda_{flight} \simeq 10^{16}m$. The other extreme example of
a Supernova with $m_{{\cal S}/{\cal P}}=1\; MeV$ ($4\; keV$) gives
$\lambda_{flight}=90\;(2 \times 10^{16})\; m$.

The above arguments make it also clear that since we have to
ensure that the particle takes away energy from the star it
cannot be produced almost at rest. The analysis becomes then a
little bit more involved as we will have to start with an
assumed value of $m_{{\cal S}/{\cal P}}$, typically $m_{{\cal
S}/{\cal P}} \simeq 10^{-1}T_{star}$, corresponding to $\displaystyle{
\left({m_{\tilde{\gamma}} \over m_{3/2}}\right) \simeq 10^{12}\left({T_{star}
\over keV}\right)}$, put the limit through exotic cooling on $m_{{\cal
S}/{\cal P}}$ (alternatively on
$\displaystyle{\left({m_{\tilde{\gamma}} \over m_{3/2}}\right)}$) and
finally check if the bound so obtained is better than the
starting assumption. Similar method has been used in \cite{mohanty} for
constraining the $\gamma \gamma$ coupling of a light pseudoscalar particle
from the physics of pulsars. For consistency reasons there one
starts with a mass smaller than $10^{-10}\; eV$. If
the two photon coupling depends upon mass, as it is the case for
axions in a model dependent way,  one has then
to check, from model to model, if the bound on the
particle mass so derived is better than $10^{-10}\; eV$.

The first process of interest in our case is the Primakoff reaction
$\gamma Z \to Z{\cal S},\; Z{\cal P}$ with $Z=e,\; p$. In the range
$m_{Z} >> \omega_{\gamma} >> \omega'$ one obtains
\begin{equation} \label{e8}
{\sigma_{\gamma} \over 2} \equiv \sigma (\gamma Z \to Z {\cal S})=
\sigma (\gamma Z \to Z {\cal P} )={\kappa^2 \alpha_{em} \over 6}
\left({m_{\tilde{\gamma}} \over m_{3/2}}\right)^2 \left\{2\ln \left({
\omega_{\gamma} \over m_{{\cal S}/{\cal P}}}\right) + 2\ln 2 -1 \right\}
\end{equation}
In the approximation we have used the cross section of the
Primakoff process is independent of the target mass and is the
same for spin-0 target (relevant for Helium burning stars). Note
also that the cross section becomes singular as $m_{{\cal
S}/{\cal P}} \to 0$.

The sum of the cross sections for all possible $\gamma \gamma$-scatterings
in the limit $\omega_{\gamma} >> m_{{\cal S}/{\cal P}},\;
m_{3/2}$ reads \cite{probir2}
\begin{equation} \label{e9}
\sigma_{\gamma \gamma}=\sum_{X={\cal SS},{\cal PP}, {\cal SP}, \tilde{G}
\tilde{G}}\sigma (\gamma \gamma \to X)={7 \over 216 \pi}\kappa^4
\left({m_{\tilde{\gamma}} \over m_{3/2}}\right)^4\; \omega_{\gamma}^2
\end{equation}

The exact calculation of an energy loss ($\dot{\epsilon}$)
per unit mass and unit time
involves an integration over the initial momenta
of the squared matrix element folded with the Bose-Einstein or Fermi-Dirac
distribution and taking into account the chemical composition of
the star and possible Pauli blocking \cite{astro}.
Here we will adopt a simplified version which is
sufficient for our purposes and should give the correct
results for stars like the Sun and Red Giants. We can write the exotic cooling
rates due to the single and two photon reactions as
\begin{equation} \label{e10}
\dot{\epsilon}_{\gamma}={n_e n_{\gamma} \sigma_{\gamma} \omega'
\over \varrho}\;\;\; + \;\;\; {n_p n_{\gamma} \sigma_{\gamma} \omega'
\over \varrho} \end{equation}
\begin{equation} \label{e11}
\dot{\epsilon}_{\gamma \gamma}={2 n_{\gamma}^2 \sigma_{\gamma \gamma}
\omega' \over \varrho} \end{equation}
where $\varrho$ is the core density of the star and $n_X$ is the
number density of the particle species $\gamma ,\; e, \; p$. The
factor $2$ appears in eq.(11) because two particles are emitted.
With $\omega_{\gamma} \sim \omega' \sim T$ and $\varrho \simeq
n_p m_N$ ($m_N$ is the nucleon mass) one obtains
\begin{eqnarray}
\label{e12}
\dot{\epsilon}_{\gamma} &\simeq & { 4\zeta (3) \over 3 \pi^2} \alpha_{em}
{\kappa^2 \over m_N} \left({m_{\tilde{\gamma}} \over m_{3/2}}\right)^2
T^4 \left\{2\ln \left({T \over m_{{\cal S}/{\cal P}}}\right) +
2\ln 2 -1 \right\} \nonumber \\
& \sim & {\zeta (3) \over 27 \pi^2}\alpha_{em}{m_{{\cal S}/{\cal P}}^2
\over \Lambda_{QCD}^4m_N} T^4 \left\{2\ln \left({T \over
m_{{\cal S}/{\cal P}}}\right) + 2\ln 2 -1 \right\}
\end{eqnarray}
\begin{equation}
\label{e13}
\dot{\epsilon}_{\gamma \gamma}\simeq {7 \over 27}{\zeta^2 (3)
\over \pi^5}{T^9 \over \varrho}\kappa^4
\left({m_{\tilde{\gamma}} \over m_{3/2}}\right)^4 \sim {7 \over
34992}{\zeta^2 (3) \over \pi^5}{T^9 m_{{\cal S}/{\cal P}}^4 \over
\varrho \Lambda_{QCD}^4}
\end{equation}
where $\zeta$ is Riemann's zeta function.

The bound on $\displaystyle{\left({m_{\tilde{\gamma}} \over
m_{3/2}}\right)}$ can be obtained now from eqs.(12) and (13) by imposing
$\dot{\epsilon}_{\gamma},\; \dot{\epsilon}_{\gamma \gamma} <
\dot{\epsilon}_{star}$. The set of astrophysical parameters
needed in the analysis are as follows: $T_{\odot} \simeq 1\; keV$,
$\varrho_{\odot}
\simeq 10^2g\; cm^{-3}$, $\dot{\epsilon}_{\odot} \simeq 17.5\; erg \;
g^{-1}\; sec^{-1}$ and $T_{RG} \simeq 10 \; keV$,
$\varrho_{RG} \simeq 10^4 \; g \; cm^{-3}$,
$\dot{\epsilon}_{RG} \simeq 10^2 erg \; g^{-1}\; sec^{-1}$. As
explained above one assumes a starting value, $m_{{\cal S}/{\cal
P}} \simeq 10^{-1}T_{star}$ or equivalently
$\displaystyle{\left({m_{\tilde{\gamma}} \over m_{3/2}}\right) \simeq
10^{12}\left({T_{star} \over keV}\right)}$, which is consistent with
temperature of the Sun and the decay lifetime of the particle.
The bound we obtain this way from eq.(12) is for the Sun $\displaystyle{
\left({m_{\tilde{\gamma}} \over m_{3/2}}\right)_{\odot} < 2.4
\times 10^9}$ and for the Red Giants $\displaystyle{
\left({m_{\tilde{\gamma}} \over m_{3/2}}\right)_{RG} < 6
\times 10^7}$. Indeed the analysis is consistent since the bound
so obtained is orders of magnitude stronger than the starting
value. One can also perform a slightly better analysis by an
iteration process. Translating the bound on $\displaystyle{
\left({m_{\tilde{\gamma}} \over m_{3/2}}\right)}$ into the
corresponding value of $m_{{\cal S}/{\cal P}}$ one can insert
the latter back into eq.(12) and proceed till a convergence is reached.
Note that the convergence is then independent whether one starts
with $m_{{\cal S}/{\cal P}} \simeq 10^{-1}T_{star}$ or
$m_{{\cal S}/{\cal P}} \simeq 10^{-2}T_{star}$. As a check one
can use the second form of eq.(12) expressed in $m_{{\cal S}/{\cal P}}$.
The conclusion is then that the following region is excluded
\begin{equation} \label{e14}
1.8 \times 10^9\;\;\; \vert \!\!\!\!\! <\;\;\;
\left({m_{\tilde{\gamma}} \over m_{3/2}}\right)_{\odot}\;\;\; \vert \!\!\!\!\!
<\;\;\; 10^{12}, \;\;\;\;\;
3.3 \times 10^7\;\;\; \vert \!\!\!\!\! <\;\;\;
\left({m_{\tilde{\gamma}} \over m_{3/2}}\right)_{RG}\;\;\; \vert \!\!\!\!\!
<\;\;\; 10^{13}
\end{equation}
These limits justify our analysis as it excludes huge range of
the parameter space. On the other hand in view of the first (second)
bound in eq.(4) it also means that the range of
$\displaystyle{
\left({m_{\tilde{\gamma}} \over m_{3/2}}\right)}$ between
$10^{13}$ and $1.8 \times 10^{15}$ ($10^{13}$ and $4.5 \times
10^{13}$) remains unrestricted by our analysis. It is worth
emphasizing that taking the bound from $p\bar{p}$ colliders
and combining it with eq.(14) this window of allowed values
becomes quite narrow.

Let us also make a rough estimate of what is to be expected for
a White Dwarf ($T_{WD} \simeq 2\; keV$, $\varrho_{WD} \simeq 10^5
g\; cm^{-3}$ , $\dot{\epsilon}_{WD} \simeq 5 \times 10^{-3} erg~
g^{-1}sec^{-1}$). Here the plasmon effects play a non-negligible
role and can be taken into account approximately by multiplying
$\dot{\epsilon}_{\gamma}$ by the
suppression factor $e^{-m_* /T}$ where $m_*$ is the
plasmon mass. As compared to the Sun the gain/loss factor is
$\displaystyle{\left[\left({T_{WD} \over T_{\odot}}\right)^4 \left({
\dot{\epsilon}_{\odot} \over \dot{\epsilon}_{WD}}\right)
e^{-m_*/T_{WD}}\right]^{-1/2} \sim  1.2 \times 10^{-2}}$ for $m_*/T_{WD}
\sim 2$. Roughly speaking the bound from White Dwarfs is not
expected to be better than the one obtained already for Red Giants.

The two gamma cooling rate is, as one can see from eq.(13), very
sensitive to the temperature. However, even for stars with high core
temperature this cannot overcome the other small factors, like $\kappa^4$,
to make the rate appreciable. For the Sun the limit one gets
from eq.(13) is $m_{{\cal S}/{\cal P}} < 3 \times 10^5\; keV$ which
is clearly not a useful bound at all. For a Supernova ($T_{SN} \simeq 30\;
MeV$, $\varrho_{SN}\simeq 10^{12}g\;cm^{-3}$, $\dot{\epsilon}_{SN}\simeq
3.6 \times 10^{19}g\; cm^{-3}$) the gain loss factor is
$\displaystyle{\left[\left({T_{SN} \over T_{\odot}}\right)^9 \left({
\varrho_{\odot} \over \varrho_{SN}}\right)\left({\dot{\epsilon}_{\odot}
\over \dot{\epsilon}_{SN}}\right)
e^{-2m_*/T_{SN}}\right]^{-1/4} \sim  10^{-2}}$ for $m_*/T_{WD}
\sim 5$. This gives $m_{{\cal S}/{\cal P}} < 1\; MeV$ which is
much weaker than the collider bound. Thus the Supernova analysis
is not likely to provide a better bound than in eq.(14) (this is
true for the processes we have investigated here, an additional process
relevant for Supernova could be $e^+e^- \to \gamma^* \to \gamma
+ {\cal S}/{\cal P}$).

In conclusion, we have derived a bound on the ratio
$\displaystyle{
\left({m_{\tilde{\gamma}} \over m_{3/2}}\right)}$ which excludes
values over several orders of magnitude. The window around $10^{12}-
10^{13}$ which our analysis leaves unconstrained can probably be
closed in future collider experiments. In this case the absolute
bound $\displaystyle{
\left({m_{\tilde{\gamma}} \over m_{3/2}}\right) < 3.3
\times 10^7}$ could be put. With $m_{\tilde{g}} > 140\;GeV$
\cite{abe} this
on the other hand would imply $m_{3/2} > 0.7 \; keV$ which comes
close to the bound from cosmology (eq.(5)) and hence leaves
only a small window for light gravitino. Note that the inherent
astrophysical uncertainties might introduce a certain margin of fluctuation
of the numbers stated above, but the main conclusions remain unchanged.

\vglue 1cm

{\elevenbf \noindent Acknowledgments \hfil}
\vglue 0.4cm
We thank A.~S.~Joshipura, P.~Roy and R.~M.~Godbole for valuable discussions.
M.N. wishes to thank the Alexander von Humboldt foundation for
financial support under the Feodor-Lynen Fellowship program.


\newpage
\begin{thebibliography}{99}
\bibitem{susy}
H.~E.~Haber and G.~L.~ Kane, Phys.~Rep.~{\bf 117}
(1985) 75; for recent review see X.~Tata in {\it The Standard
Model and Beyond} p.~304, ed.~J.~E.~Kim, World Scientific 1991
\bibitem{gut}
G.~G.~Ross, {\it Grand Unified Theories}, The Benjamin/Cummings
Publishing Company 1985 and references therein
\bibitem{sugra}
P.~Nieuwenhuizen, Phys.~Rep.~{\bf 68} (1981) 189; P.~Nath,
R.~Arnowitt and A.~Chamseddine, {\it Applied N=1 Supergravity},
ICTP Series in Theoretical Physics, Vol.~1, World Scientific 1984
\bibitem{pdb}
{\it Review of Particle Properties}, Phys.~Rev~{\bf D45} (1992)
\bibitem{fayet1}
P.~Fayet, Phys.~Lett.~{\bf B70} (1977) 461
\bibitem{ellis1}
J.~Ellis, K.~Enqvist and D.~V.~Nanopoulos, Phys.~Lett.~{\bf
B147} (1984) 99; for a review see A.~B.~Lahanas and D.~V.~Nanopoulos,
Phys.~Rep.~{\bf 145} (1987) 1
\bibitem{probir1}
T.~Bhattacharaya and P.~Roy, Phys.~Lett.~{\bf B206} (1988) 655;
Nucl.~Phys.~{\bf B328} (1989) 469; {\it ibid} {\bf B328} (1989) 481
\bibitem{fayet2}
P.~Fayet, Phys.~Lett.~{\bf B175} (1986) 471
\bibitem{probir2}
T.~Bhattacharaya and P.~Roy, Phys.~Rev.~Lett.~{\bf 38} (1987)
1517; Phys.~Rev.~{\bf D38} (1988) 2284
\bibitem{nandi}
D.~A.~Dicus, S.~Nandi and J.~Woodside, Phys.~Rev.~{\bf D41}
(1990) 2347
\bibitem{dicus}
D.~A.~Dicus and P.~Roy, Phys.~Rev.~{\bf D42} (1990) 938
\bibitem{primack}
H.~Pagels and J.~R.~Primack, Phys.~Rev.~Lett.~{\bf 48} (1982) 223
\bibitem{weinberg}
S.~Weinberg, Phys.~Rev.~Lett.~{\bf 48} (1982) 1303
\bibitem{cremmer}
E.~Cremmer, S.~Ferrara, L.~Girardello and A.~van Proyen,
Nucl.~Phys.~{\bf B212} (1983) 413
\bibitem{ellis2}
J.~Ellis, K.~Enqvist and D.~V.~Nanopoulos, Phys.~Lett.~{\bf
B151} (1985) 357
\bibitem{nilles}
H.~P.~Nilles, Phys.~Rep.~{\bf 110} (1984) 1 and references therein
\bibitem{ellis3}
S.~Dimopoulos and S.~Raby, Nucl.~phys.~{\bf B219} (1983) 479;
J.~Ellis, A.~D.~Linde and D.~V.~Nanopoulos, Phys.~Lett.~{\bf
B118} (1982) 59
\bibitem{linde}
M.~Y.~Khlopov and A.~D.~Linde, Phys.~Lett.~{\bf B138} (1984) 265
\bibitem{cohen}
For a recent review and further references see A.~G.~Cohen,
D.~B.~Kaplan and A.~E.~Nelson,
Annu.~Rev.~Nucl.~Part.~Sci.~{\bf 43} (1993) 27.
\bibitem{kuzmin}
V.~A.~Kuzmin, V.~A.~Rubakov and M.~E.~Shaposhnikov,
Phys.~Lett.~{\bf B155} (1985) 36; M.~Fukugita and T.~Yanagida,
Phys.~Lett.~{\bf B174} (1986) 45;
M.~A.~Luty, Phys.~Rev.~{\bf D45} (1992) 455; A.~Acker et al.,
Phys.~Rev.~{\bf D48} (1993) 5006
\bibitem{rparity}
V.~S.~Berezinsky, Phys.~Lett.~{\bf B261} (1991) 71;
J.~A.~Frieman and G.~F.~Giudice, Phys.~Lett.~{\bf B224} (1989) 125
\bibitem{astro}
For processes used in astrophysics to restrict the axion mass see
M.~Fukugita, S.~Watamura and M.~Yoshimura, Phys.~Rev.~{\bf D26}
(1982) 1840; A.~Pantziris and K.~Kang, Phys.~Rev.~{\bf D33}
(1986) 3509 and earlier references therein.
\bibitem{note1}
Most of the astrophysical values are taken from ref. [22].
The energy loss of the Sun at the center $\epsilon_{\odot} \simeq
17.5\;erg\;g^{-1}sec^{-1}$ is taken from D.~D.~Clayton, {\it
Principles of Stellar Evolution and Nucleosynthesis}, University
of Chicago Press 1983
\bibitem{bethe}
H.~A.~Bethe, Rev.~Mod.~Phys.~{\bf 62} (1990) 801
\bibitem{mohanty}
S.~Mohanty and S.~N.~Nayak, Phys.~Rev.~Lett.~{\bf 70} (1993) 4038
\bibitem{abe}
F.~Abe et al., Phys.~Rev.~Lett.~{\bf 69} (1992) 3439
\end{thebibliography}
\end{document}